# Analysis of BOLD fMRI Signal Preprocessing Pipeline on Different Datasets while Reducing False Positive Rates

Yunxiang Ge [a, +], Yu Pan [b, +], Weibei Dou [a, *]

[a] Department of Electronic Engineering, Tsinghua University, Beijing 100084, China

[b] Department of Rehabilitation, Tsinghua Changgung Hospital, Beijing 102218, China

[+] The two authors contributed equally to this paper.

[*] Corresponding author at: Department of Electronic Engineering, Tsinghua University, Beijing 100084, China. Tel.: +86-10-62781703; E-mail address: douwb@tsinghua.edu.cn

## Abstract

The technology of functional Magnetic Resonance Imaging (fMRI) based on Blood Oxygen Level Dependent (BOLD) signal has been widely used in clinical treatments and brain function researches. The BOLD signal has to be preprocessed before being analyzed using either functional connectivity measurements or statistical methods. Current researches show that data preprocessing steps may influence the results of analysis, yet there is no consensus on preprocessing method. In this paper, an evaluation method is proposed for analyzing the preprocessing pipeline of resting state BOLD fMRI (rs-BOLD fMRI) data under putative task experiment designs to cast some lights on the preprocessing stage, covering both first and second level analysis. The choices of preprocessing parameters and steps are altered to investigate preprocessing pipelines while observing statistical analysis results, trying to reduce false positives as reported by Eklund et al. in their 2016 PNAS paper. All of the experiment data are separated into 7 datasets, consisting of 220 healthy control samples and 136 patient data that are from 38 incomplete Spinal Cord Injury (SCI) patients and 16 Cerebral Stroke (CS) patients, including multiple scans of some patients at different time. These data were acquired from two different MRI scanners, which may cause difference in analysis results. The evaluation result shows that it has little effect to change parameters in each steps of the classical preprocessing pipeline, which consists head motion correction, normalization and smoothing. Removing time points and the following detrend step can reduce false positives. However, covariates regression and filtering has complicated effects on the data. Note that for single subject analysis, false positives declined consistently after filtering. The result of patient data and healthy controls data which are scanned under the same machine with the same acquisition protocol shows little difference. Yet data acquired with different scanner and protocol influences statistical analysis significantly. As a result, future research should pay more attention on the scanning machine and protocol used. This research is a preliminary investigation on rs-BOLD fMRI signal preprocessing. We hope the conclusions can be of value for other studies in the field of brain function research.

Key words: Blood Oxygen Level Dependent signal; functional Magnetic Resonance Imaging; preprocessing; statistics; false positive rates

## 1 Introduction

The technology of functional Magnetic Resonance Imaging (fMRI) has been widely used in clinical diagnosis and neuroplasticity researches. It has the benefit of relative high spatial and temporal resolution, and thus is suitable to investigate the functional connectivity of human brain. Since resting state fMRI (rs-fMRI) does not require the subject to perform any specific task, it is often used in hospitals where patients are not able to follow complicated instructions. Also, rs-fMRI is becoming popular in neuroplasticity studies [1]. During a scanning session, fMRI directly measures changes in the Blood Oxygen Level Dependent (BOLD) signal, reflecting neural activities. The acquired data contain imperfections and artifacts due to subject movement, spontaneous neuro activities and intrinsic electron thermal noises. As a result, certain degree of preprocessing must be performed before analyzing the data using either functional connectivity measurements or statistical methods.

The preprocessing of fMRI data is performed step by step, forming the preprocessing pipeline. Currently there is no consensus on what steps should be performed in the preprocessing stage, or how to choose parameters in those steps. Yet preprocessing poses huge impact on the BOLD fMRI data, and there have been many studies on this issue. Stephen C Strother [2], for example, analyzed common preprocessing steps for BOLD fMRI and their possible influences. Jonathan D Power et al. [3] found that without proper preprocessing, subject motion can cause spurious but systematic correlations in functional connectivity MRI networks. The signal changes induced by motion increase observed resting state functional connectivity (RSFC) [4]. Changwei W. Wu et al. did an empirical study on how slice-timing, smoothing and normalization affect seed-based rs-fMRI correlation analysis [5]. Ronald Saldky et al., on the other hand, evaluated the impact of the slice-timing effect on simulated data for different fMRI paradigms and measurement parameters, emphasizing the significance of slice-timing correction methods [6]. Michael

N Hallquist et al. [7] found that exchanging nuisance regression and filtering in the preprocessing pipeline would produce different results. William R. Shirer et al. [8] tried to identify the data preprocessing pipeline that optimizes rs-fMRI data across multiple outcome measures, such as signal-to-noise ratio, test-retest reliability and group discriminability. All of the above studies show that preprocessing influences BOLD fMRI data analysis in many ways.

In 2016, Eklund et al. [9] analyzed several rs-fMRI data of healthy controls using task related statistical methods and found inflated false-positive rates for clusterwise inference. Guillaume Flandin and Karl J. Friston, on the other hand, emphasized the advantages of parametric analyses in their technical report [10]. We noticed that Eklund et al. used the default preprocessing parameters of 3 different processing packages in that work. It was, however, not clear how different preprocessing steps may influence the false-positive rates. On the other hand, the subjects involved in that work are healthy controls, while neuroplasticity researches are also concerned with patients of different kinds of neural disabilities. What kind of preprocessing should researchers adopt for patient data?

In this paper, we try to expand the research by using rs-fMRI data of 2 kinds of patients, as well as healthy controls. The patients have either incomplete Spinal Cord Injury (SCI) or Cerebral Stroke (CS), but not both. Part of the healthy controls data are scanned from volunteers and the others are downloaded from the 1000 functional Connectomes Project [11] which is also used as an online data sharing center by Eklund et al. [9]. Since the sample size of each dataset is not the same, we evaluate the possible influence of this by conducting experiments on different scales of one dataset. Besides, we try to appreciate the influence of scanning machines in our present work. Specifically, an analyzing framework suitable for our goal is developed. We evaluate the preprocessing pipeline using different datasets by changing preprocessing steps and parameters, while observing statistical analysis results after each major step.

We find that changing parameters in certain preprocessing steps has little influence on the result. Expanding the preprocessing pipeline by adding certain steps may reduce false positives. The patient data and healthy controls data which are scanned under the same machine shows little difference. On the other hand, the source of the data influences statistical analysis, which implies that it is necessary to use different preprocessing methods for data acquired from different scanning machines.

The structure of this paper is as follows. We explain the analyzing method in the second section, where the analyzing framework is firstly introduced in detail. All of the preprocessing methods used in our work are described in this section, followed by a thorough explanation of all datasets used, including details about the scanning process. Then the statistical analysis method is briefly reviewed. In the third section, we present our main experiment results using tables and curves. And lastly, some points and questions are discussed in the fourth section.

## 2 Method

The procedure of this study follows the work of Eklund et al. [9], where resting-state fMRI data are analyzed with putative task designs. Unlike their work, we use DPARSFA[1] (Data Processing Assistant for Resting-State fMRI) [12] to preprocess data. Some of the preprocessing steps in DPARSFA are the same as in SPM[2] (Statistical Parametric Mapping) [13, 14]. After preprocessing, further analysis is performed using SPM 12. Only one of the four task experiment designs used by Eklund et al. [9] is considered in our work. Later, both first level analysis (for a single subject) and second level analysis (for a group of subjects) is performed independently. We use the MATLAB scripts provided by Eklund et al. [9] from their Github repository[3] to perform statistical analysis.

As Eklund et al. reported in [9], parametric methods used in SPM can give a very high degree of false positives for clusterwise inference. In this study, we fix the statistical analysis procedure, simply using clusterwise inference alone and try to reduce the false positives by changing preprocessing steps and parameters for healthy controls. Later, we include patient data into the experiment, keeping the whole program pipeline unchanged (**Figure 1**).

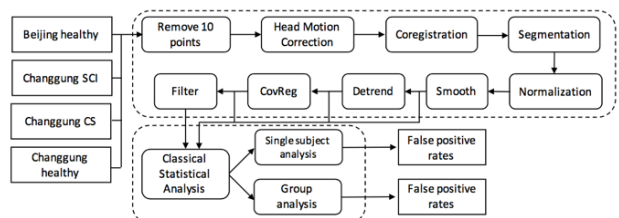

**Figure 1.** The whole program pipeline. All datasets used are listed at the left side of the figure. Upper dotted bounding box represents preprocessing steps. Statistical analysis is surrounded in the lower bounding box. The analysis results are false positive rates. The arrow keys show data flow directions. Notice there are multiple inputs to the statistical analysis program from preprocessing stage.

### 2.1 Preprocessing

All of the preprocessing steps used in this study are integrated in the DPARSFA toolbox [12], including removing time points, realignment, coregistration, segmentation, normalization, smoothing, detrending, covariates regression and filtering. DPARSFA also saves data after each step, enabling comparative study of these steps. Slice timing correction is not performed, since slice time information is insufficient. Coregistration and segmentation are performed only when the corresponding T1 image is available. For normalization, we use Echo-Planar Imaging (EPI) template [12, 15]. Long-term physiological shifts, movement related noise remaining after realignment or instrumental instability may contribute to a systematic increase or decrease in the signal with time [12, 16, 17]. We use a linear model as does in Analysis of Functional Neuroimage (AFNI) [18] for detrending [12]. The white matter signals, the cerebrospinal fluid (CSF) are removed as nuisance variables. Also, Global Signal Regression (GSR) is performed. The data is then filtered to reduce influence of respiratory and aliased cardiac signals [12, 19].

---

[1] http://rfmri.org/DPARSF, release V4.3
[2] http://www.fil.ion.ucl.ac.uk/spm/
[3] https://github.com/wanderine/ParametricMultisubjectfMRI

Some major steps have a capital letter abbreviation code for easy reference (**Table 1**). In our work, we choose different preprocessing pipelines as described below.

**Table 1**. Major preprocessing steps and their abbreviation codes.

| Code | Step |
|---|---|
| R | Realign |
| W | Normalization |
| S | Smooth |
| D | Detrend |
| C | Covariates Regression |
| F | Filtering |

The most common one is called the classical preprocessing pipeline. It simulates task based fMRI data analysis, taking BOLD fMRI data and T1 image as input. The fMRI data is firstly motion corrected, then coregistered with structural image. Segmentation and skull extraction is performed before normalizing to the same brain atlas. The last step is spatial smoothing with 4 different Full Widths Half Maximum (FWHM) Gauss kernel. We use 4mm, 6mm, 8mm and 10mm FWHM in our work. Later on we keep FWHM of the smoothing step to 6mm, since the false positive rates are the highest [9]. For other steps of the preprocessing pipeline, parameters are kept at default. Based on the step choice, the classical preprocessing pipeline is abbreviated as RWS (See **Table 1**).

We also investigate how parameter choices in the classical preprocessing pipeline influence analyze results by changing parameters in each step, keeping smoothing FWHM at 6mm. The parameters in other steps are kept at default while we change parameters at one step. The parameter choices are summarized in **Table 2**.

Further, we extend the classical preprocessing pipeline by removing 10 time points from the data, and append detrending, covariates regression and filtering steps to the pipeline. It is thus called the 'extended preprocessing pipeline'. The pass band of the filter is set to 0.01-0.08Hz according to [12]. Also, the FWHM of smoothing kernel is kept at 6mm. In order to evaluate the effect of newly added steps, data are retrieved after each steps and fed into statistical analysis program separately. The resulting pipelines are abbreviated as RWSD, RWSDC, RWSDCF and so on (See **Table 1**).

## 2.2 Data

Our present work uses data of both healthy controls and patients. For healthy control data, we downloaded 198 subjects from the 1000 Functional Connectomes Project [20] and named them as the Beijing dataset. The data was acquired using a Siemens TRIO TIM scanner with 3.0T magnetic strength [21]. We also include data from 22 healthy volunteers scanned at Beijing Tsinghua Changgung hospital, using a GE DISCOVERY MR750 scanner with 3.0T magnetic strength. Other parameters of the scanning are summarized in **Table 3**. In order to avoid the influence of sample size, we randomly choose 38 subjects from the Beijing dataset to form a smaller one, called Beijing r38 dataset.

**Table 2**. Selected steps in the classical preprocessing pipeline and their parameter changes.

| Step | Algorithm | Parameter | Range | Default |
|---|---|---|---|---|
| Realign | Realign: Estimate & Reslice | Estimate option: Separation | 1-6 | 4 |
|  |  | Estimate option: Interpolation | 1-7 | 2 |
|  |  | Reslice option: Interpolation | 0-7 | 4 |
| Coregistration | Coregister: Estimate | Objective Function | 1-3 | 2 |
| Segmentation | Old Segment | Sampling distance | 1-5 | 3 |
| Normalization | Old Normalise: Write | Interpolation | 0-7 | 1 |

**Table 3**. Scanning details comparison between Beijing and Changgung dataset.

| Parameters | Beijing | Changgung |
|---|---|---|
| Field Strength/T | 3.0 | 3.0 |
| Scanner | Siemens TRIO TIM | GE DISCOVERY MR750 |
| Scan series | Gradient echo EPI | EP, GR |
| TR/s | 2.0 | 2.0 |
| TE/ms | 30 | 30 |
| #slices | 33 | 34 |
| Matrix size | 64×64 | 64×64 |
| FOV/mm | 240 | 224 |
| Voxel size/mm | 3.75×3.75×3.50 | 3.50×3.50×3.50 |
| #volumes | 225 | 240 |

The patient data used in this study comes from the department of rehabilitation, Beijing Tsinghua Changgung hospital, including 38 incomplete Spinal Cord Injury (SCI) patients and 16 Cerebral Stroke (CS) patients. Informed consents were obtained from all patients. For each SCI patient, there are one to six scans at different time, totaling 87 samples. The total number of stroke patient data sample is 49. Here we investigate data from two types of disease separately, and split data from the same type of disease into two datasets, the full dataset and the single dataset. So we have Changgung SCI dataset with 87 SCI patient samples, including multiple scans for the same person, Changgung SCI single dataset with 38 SCI patient samples, including single scans for the same person, and the same for cerebral stroke patients. All datasets used in the current study are summarized in **Table 4**. Sample images of raw BOLD fMRI data are shown in **Figure 2**.

**Table 4**. Datasets used in this study.

| Dataset | Size | Subject category | Note |
|---|---|---|---|
| Beijing | 198 | Healthy | |
| Beijing r38 | 38 | Healthy | Randomly selected |
| Changgung SCI | 87 | Spinal Cord Injury | Multiple scans |
| Changgung SCI single | 38 | Spinal Cord Injury | Single scan |
| Changgung CS | 49 | Cerebral Stroke | Multiple scans |
| Changgung CS single | 16 | Cerebral Stroke | Single scan |
| Changgung normal | 22 | Healthy | |

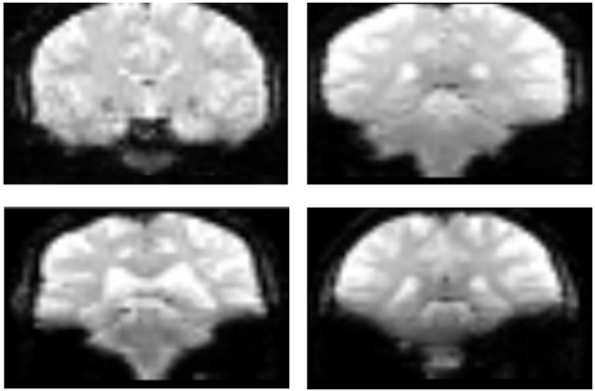

**Figure 2**. Raw BOLD fMRI sample images for 4 different datasets. Only images at the first time point are shown here. Top left: Beijing, top right: Changgung SCI, bottom left: Changgung CS, bottom right: Changgung normal.

### 2.3 Statistic methods

After data preprocessing, we perform classical statistical analysis and first, second level analysis. According to [9], we only use one putative task design in this work. The task design is called 'Event1' in this paper, which consists 2 seconds activation and 6 seconds rest. Because the resting-state fMRI data should contain no consistent shifts in BOLD activity, for a group of subjects the null hypothesis of mean zero activation should be true [9]. For a single subject, resting-state fMRI data should not contain any activation that correlates with the experiment design [22]. The result of first and second level analysis, which is the false positive rates, can be calculated as the proportion of analyses that give rise to significant results [9]. We use cluster-wise inference in this work, and setting Cluster Defining Threshold, also known as CDT, to p=0.01 (CDT1) and p=0.001 (CDT2). The results, which are Family-wise Error rates, are denoted as FWE1 and FWE2 respectively. Second level analyses are repeated 1000 times for each preprocessing and experiment settings. Two group sizes, 10 and 40, are used depending on the dataset.

All datasets used are summarized in **Table 4**. For the Beijing dataset, we set group size to 40 to match the settings in [9], in order to investigate influences of parameter changes. For the rest datasets, group size is kept at 10, since there are not enough subjects in some datasets for a 40 people group. Also, almost all single subject datasets from Changgung hospital consist small amount of subjects compared with Beijing dataset. We perform analysis on a randomly selected subset of Beijing dataset for compare. Results of all experiments are compared between different datasets and different preprocessing steps.

## 3 Result

### 3.1 Results of classical preprocessing pipeline

37 experiments are performed on Beijing dataset, using different parameter choices based on the classical preprocessing pipeline. The results are summarized in **Table 5**.

**Table 5**. Results of classical preprocessing pipeline.

| Parameters | Value | Second level | | First level | |
|---|---|---|---|---|---|
| | | FWE1/% | FWE2/% | FWE1/% | FWE2/% |
| Realign_est_sep | 1 | 57.9 | 27.6 | 18.69 | 8.08 |
| | 2 | 57.9 | 27.6 | 18.69 | 8.08 |
| | 3 | 57.1 | 26.7 | 19.19 | 8.08 |
| | 4 | 59.3 | 27.9 | 18.69 | 7.58 |
| | 5 | 57.7 | 28.2 | 19.19 | 6.57 |
| | 6 | 58.3 | 28.5 | 20.71 | 7.07 |
| Realign_est_interp | 1 | 58.1 | 27.9 | 19.70 | 6.57 |
| | 2 | 59.3 | 27.9 | 19.70 | 7.58 |
| | 3 | 59.0 | 28.1 | 18.69 | 7.58 |
| | 4 | 59.0 | 28.3 | 19.19 | 8.08 |
| | 5 | 57.7 | 27.8 | 19.19 | 7.58 |
| | 6 | 57.9 | 28.9 | 19.19 | 6.57 |
| | 7 | 57.4 | 28.0 | 18.18 | 6.57 |
| Realign_res_interp | 0 | 59.7 | 29.1 | 19.19 | 7.07 |
| | 1 | 59.1 | 28.8 | 19.19 | 8.08 |
| | 2 | 58.9 | 28.7 | 18.69 | 7.58 |
| | 3 | 59.1 | 28.9 | 18.69 | 8.08 |
| | 4 | 59.3 | 27.9 | 19.7 | 7.58 |
| | 5 | 59.4 | 28.9 | 18.69 | 7.58 |
| | 6 | 58.8 | 28.2 | 18.18 | 7.58 |
| | 7 | 58.7 | 28.8 | 18.69 | 8.08 |
| Coregister_est | mi | 59.0 | 28.3 | 18.18 | 8.08 |
| | nmi | 59.3 | 27.9 | 19.70 | 7.58 |
| | ecc | 58.4 | 28.5 | 19.19 | 8.08 |
| Oldseg_sampd | 1 | / | / | / | / |
| | 2 | 58.3 | 28.4 | 18.69 | 7.58 |
| | 3 | 59.3 | 27.9 | 19.70 | 7.58 |
| | 4 | 58.5 | 28.8 | 18.69 | 8.08 |
| | 5 | 59.7 | 29.1 | 18.18 | 8.08 |
| Oldnorm_interp | 0 | 56.4 | 26.6 | 18.69 | 5.05 |
| | 1 | 59.3 | 27.9 | 19.70 | 7.58 |
| | 2 | 57.3 | 25.5 | 19.70 | 7.58 |
| | 3 | 57.2 | 25.9 | 19.70 | 8.08 |
| | 4 | 56.7 | 25.5 | 20.20 | 6.57 |
| | 5 | 57.2 | 25.8 | 19.70 | 7.58 |
| | 6 | 57.0 | 25.9 | 18.18 | 8.08 |
| | 7 | 57.1 | 25.6 | 19.19 | 8.08 |

The cells with yellow background represent the default setting. Experiment for Oldseg_sampd=1 was not performed due to a program bug. But from the table it is easy to see that the results hardly vary, for either first level or second level analysis. The maximum change is below 3%. Also, the result does not change in any fixed pattern, either growing or decreasing as the related parameter changes. We conclude that changing parameters in the classical preprocessing pipeline has little influence on statistical analysis results.

## 3.2 Results of extended preprocessing pipeline

We conduct several experiments using the extended preprocessing pipeline on Beijing dataset (**Table 6**).

**Table 6**. Results on Beijing dataset using extended preprocessing pipeline. The first column shows preprocessing method used, where the first row acts like a ground truth.

| Method | Steps | Second level | | First level | |
|---|---|---|---|---|---|
| | | FWE1/% | FWE2/% | FWE1/% | FWE2/% |
| / | (baseline) | 59.3 | 27.9 | 19.7 | 7.6 |
| default | RWS | 59.7 | 28.5 | 20.7 | 7.58 |
| | RWSD | 40.7 | 14.9 | 18.69 | 8.08 |
| | RWSDC | 37.0 | 15.4 | 17.17 | 5.56 |
| | RWSDCF | 29.1 | 8.4 | 0 | 0 |
| | gRWSDC | 31.3 | 12 | 16.16 | 6.06 |
| | gRWSDCF | 33.9 | 10.5 | 0 | 0 |
| nocov | RWSD | 40.7 | 14.9 | 18.69 | 8.08 |
| | RWSDF | 47.2 | 19.7 | 0.5 | 0.5 |
| rm10ptn | RWS | 26.4 | 9.2 | 15.15 | 5.56 |
| | RWSD | 8.3 | 2.4 | 12.12 | 5.05 |
| | RWSDC | 24.5 | 11.1 | 17.17 | 6.57 |
| | RWSDCF | 15.2 | 3.4 | 0.05 | 0.05 |
| | gRWSDC | 31.5 | 11 | 18.18 | 10.10 |
| | gRWSDCF | 21.2 | 4.8 | 0 | 0 |

Here rs-fMRI data is analyzed without the corresponding T1 image. The coregistration, segmentation step is not performed as a result. For the default method, all parameters are kept at their default value. The capital letters in the steps column represent preprocessing steps performed. Note that the 'g' prefix means Global Signal Regression (GSR) during nuisance covariates regression. The nocov method means 'no covariates regression', where the settings of previous steps are the same as in the default setting so they are not listed in the table. The 'rm10ptn' method means removing first 10 time points from the data, which might influence the fMRI time series. We listed results of all steps in the table. **Figure 3** is a line plot of results of second level analysis on CDT 1.

We find that removing the first 10 time points can reduce the false positive rates. The detrending step shows consistent effects under different experiment settings, also decreasing the false positive rates. The following covariates regression step is complicate. It can reduce false positives under some settings, yet it would cause false positives to increase. The filtering step also has confounding influence on results. For first level analysis, however, filtering can reduce false positives to zero.

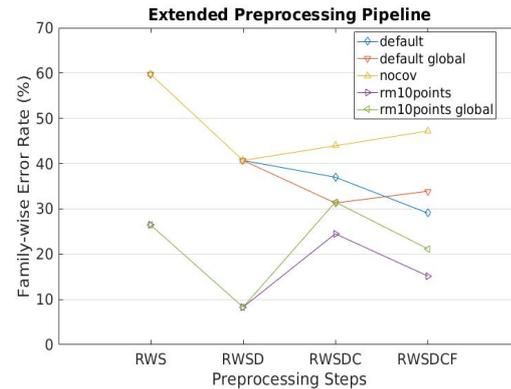

**Figure 3**. Line plot of results of second level analysis on CDT 1

## 3.3 Results of patient data

Further, we analyze patient datasets using the extended preprocessing pipeline (**Table 7**). Here we only use the default method to compare between patient datasets and between patients and Beijing healthy controls. The results of second level analysis on CDT1 are plotted in **Figure 4** and **Figure 5**.

**Table 7**. Results of Changgung patient data. Here we only use the default method to preprocess data. Each row represents results after different steps of preprocessing.

| Dataset | Steps | Second level | | First level | |
|---|---|---|---|---|---|
| | | FWE1/% | FWE2/% | FWE1/% | FWE2/% |
| SCI | RWS | 0 | 0 | 0 | 0 |
| | RWSD | 0 | 0 | 0 | 0 |
| | RWSDC | 13.2 | 4.6 | 6.9 | 1.15 |
| | RWSDCF | 45.4 | 19.1 | 0 | 0 |
| | gRWSDC | 36.5 | 12.6 | 14.94 | 5.75 |
| | gRWSDCF | 84.4 | 44.7 | 0 | 0 |
| CS | RWS | 0 | 0 | 0 | 0 |
| | RWSD | 0 | 0 | 0 | 0 |
| | RWSDC | 3 | 1 | 6.12 | 4.08 |
| | RWSDCF | 54.9 | 30.1 | 2.04 | 2.04 |
| | gRWSDC | 29.9 | 9.2 | 20.41 | 6.12 |
| | gRWSDCF | 94.7 | 65.8 | 0 | 0 |
| SCI single | RWS | 0 | 0 | 0 | 0 |
| | RWSD | 0 | 0 | 0 | 0 |
| | RWSDC | 13.2 | 4.5 | 10.53 | 2.63 |
| | RWSDCF | 54.4 | 22.3 | 0 | 0 |
| | gRWSDC | 28 | 8.8 | 7.89 | 7.89 |
| | gRWSDCF | 72.5 | 24.6 | 0 | 0 |
| CS single | RWS | 0 | 0 | 0 | 0 |
| | RWSD | 0 | 0 | 0 | 0 |
| | RWSDC | 1.9 | 0.5 | 0 | 0 |
| | RWSDCF | 37.5 | 11.7 | 0 | 0 |
| | gRWSDC | 60.3 | 18 | 12.5 | 0 |
| | gRWSDCF | 100 | 97.6 | 0 | 0 |

It can be easily noticed that second level analysis results remain zero after the classical statistical analysis pipeline (RWS), for either full dataset or single subject dataset. This is not the same as in the analysis on Beijing dataset. Also, the false positive rate after detrending is still zero. However, covariates regression and filtering increases the false positive rates significantly. Besides, we find that for two different kinds of patient, and between the two datasets of one kind of patient, the trend of statistical results is almost the same.

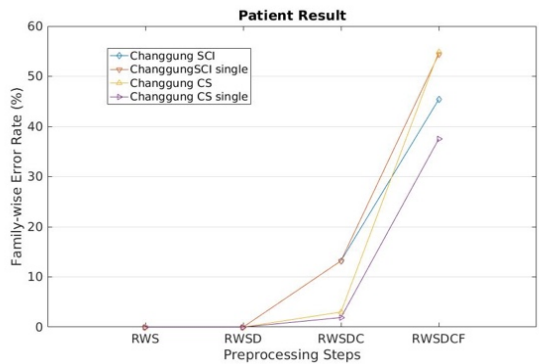

**Figure 4**. Line plot of results of second level analysis on CDT 1.

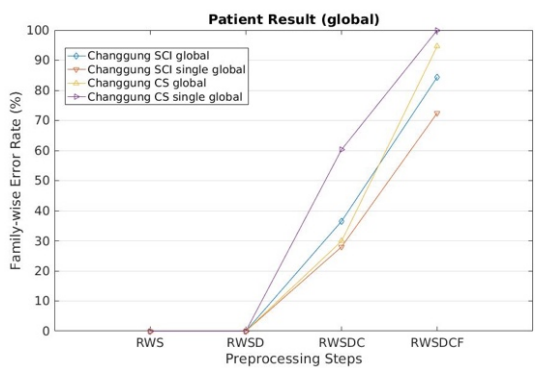

**Figure 5**. Line plot of results of second level analysis (with global signal regression) on CDT 1.

### 3.4 Results of a comparative study of two healthy datasets

In order to investigate the different result of Beijing and Changgung dataset, we conduct an experiment between Changgung normal dataset and Beijing r38 dataset (**Table 8**) and plot the result of second level analysis on CDT1 in **Figure 6**.

From the figure, the result of Beijing r38 dataset is similar with the full Beijing dataset. Differences on numbers may result from smaller samples and group sizes. And the result of Changgung normal dataset has similar trend as the results of Changgung patient datasets. This means that smaller samples or 10 people group size does not influence analysis results greatly. Also, the difference on subject body condition does not affect results. In other words, using patient data is not the cause of different results. As Beijing data and Changgung data are scanned using different machines, this may be the cause. Besides, we notice that for patient data and normal control data, first level analysis after filtering shows nearly zero false positive rates, which agrees with our previous conclusion.

**Table 8**. Results of a comparative study of two healthy datasets

| Dataset | Steps | Second level | | First level | |
|---|---|---|---|---|---|
| | | FWE1/% | FWE2/% | FWE1/% | FWE2/% |
| Changgung normal | RWS | 0 | 0 | 0 | 0 |
| | RWSD | 0 | 0 | 0 | 0 |
| | RWSDC | 34.2 | 12.9 | 0 | 0 |
| | RWSDCF | 11.4 | 2.1 | 0 | 0 |
| | gRWSDC | 15.6 | 3.9 | 9.09 | 0 |
| | gRWSDCF | 40.3 | 16.5 | 0 | 0 |
| Beijing random 38 | RWS | 69.4 | 42 | 23.68 | 13.16 |
| | RWSD | 59 | 31.5 | 23.68 | 13.16 |
| | RWSDC | 45.6 | 24.2 | 15.79 | 10.53 |
| | RWSDCF | 31 | 13.5 | 0 | 0 |
| | gRWSDC | 32.2 | 14.3 | 10.53 | 5.26 |
| | gRWSDCF | 23.5 | 8.5 | 0 | 0 |

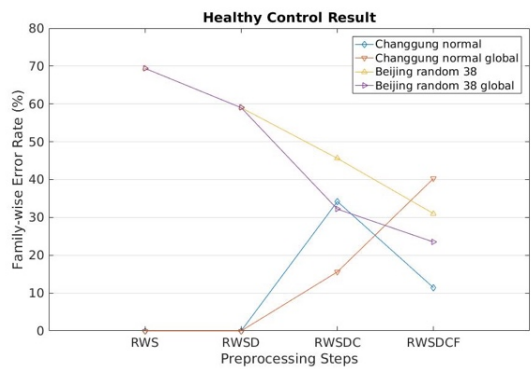

**Figure 6**. Line plot of results of second level analysis on CDT 1.

## 4 Discussion

From the above results, we conclude some advice on choosing preprocessing parameters and steps. Firstly, for the classical preprocessing pipeline which basically consists realign, normalization and smoothing, changing parameters in those steps (excluding smoothing) has little influence on the false positive rates. As for the extended preprocessing pipeline, removing time points and detrending can reduce the false positives to some extent. However, covariates regression and filtering have complicated influence. They may cause false positives to increase significantly according to our study. Besides, for first level analysis, filtering can consistently reduce false positive rates to nearly zero for all datasets.

In [12], Yan Chao-Gan and Zang Yu-Feng discussed that the first few time points in the fMRI time series are often discarded for signal equilibrium and to allow the participants' adaption to the scanning process. We find removing the first 10 time points can actually reduce false positive rates. However, it is not clear why this happens, as there is little research on effects of removing first time points except Yan and Zang's paper [12]. The detrending step, which was reported to remove non-neural signal changes in the data whose cause was not fully understood [12, 16, 17], proves to be useful in our work. Yet recently there is no paper on this issue as well. For nuisance covariates regression, it is commonly used in functional connectivity researches [7, 12, 23, 24] and less seen in classical statistical analysis. This step can cause false positives to increase dramatically according to our findings and is not recommended when performing classical statistical analysis. As for the filtering step, it is useful in single subject analysis (first level analysis) as it reduces false positives to nearly zero in almost every dataset. The filtering pass band is set to be 0.01-0.08 Hz, which excludes non-neural signal fluctuations like respiratory and cardiac artifacts [12].

We also have some findings regarding the datasets and sizes. In our work, we use multiple datasets, covering both healthy controls and 2 kinds of patients. The size of the datasets ranges from 16 to 198. For one kind of patient data, we split them into two datasets, containing one scan per patient and multiple scans per patient respectively. For each patient, a certain degree of rehabilitation treatment is performed between two consecutive scans. We thought that the existence of multiple scans for one patient could cause pitfalls in the statistical analysis step and further compromise our designed analyzing framework, since the scans of one patient may have many similarities. Our findings, however, tell a different story. The size of datasets does not play an important role. If we compare the results between Changgung SCI and Changgung SCI single (Figure 3), we can find that there is no significant variance, validating our study methodology. Also, viewing Beijing dataset and Beijing r38 dataset results can lead to the same conclusion. It shows that the statistical analysis method used in our study is robust regarding grouping sizes.

For the two kinds of patient data, some interesting conclusions can be drawn. By comparing results of one patient dataset against the other patient dataset with corresponding size, say between Changgung CS and Changgung SCI, it is clear that the results have similar trend when adding preprocessing steps (Figure 3). The data has zero false positive rates until covariates regression is performed. The following filtering step continues inflating false positives. It seems that although the neuro systems of patients show different level of malfunctions, there exist similarity in their rs-BOLD fMRI data time series. And after the same preprocessing steps, this similarity is shown by calculating the false positive rates. We can conclude that for rs-BOLD fMRI data analyzed using statistical method, the state of the subject (healthy or ill) does not affect false positive rates.

On the other hand, how may the scanning machine used influence the results? By comparing Changgung normal and Beijing r38 dataset, it is clear that the results of two datasets are not the same, despite they all consist rs-BOLD fMRI data of healthy controls. In fact, Changgung normal dataset has similar trend as other Changgung patient datasets results, while Beijing r38 dataset and Beijing dataset are alike. This further confirms that the scanner used has significant influence on statistical analysis. As a result, we advocate all future researches take a closer look at the scanner, including parameters of the scanning process and reconstruction software used by the scanner.

In the work of Eklund et al. [9], 3 datasets (Beijing, Cambridge and Oulu) are analyzed under 4 different activity paradigms, whereas here we only consider the Beijing dataset under event1 experiment design. Actually we tried to use the Cambridge dataset, but the statistical analysis program reported several errors during second level analysis, causing the false positives decrease. As a result, we fail to achieve the same results reported by Eklund et al. [9]. Also, all 4 activity paradigms were used to analyze Beijing dataset. For event2 paradigm, there are errors in the second level analysis. Among the other experiment designs, the result false positive rates of event1 is the highest without any errors. So we choose event1 as the only experiment design in our work, hoping to reduce the false positives as many as possible. During patient data analysis, we keep statistical analysis methods unchanged. So all patient data are analyzed under event1 experiment design.

For healthy control data collected at Changgung hospital, however, errors occurred again at second level analysis phase. This results in a decreased false positive rates. In other words, the actual false positives should be no less than the calculated value reported in Table 8. Yet from Figure 5, we can see that the curve related to healthy data at Changgung shows similar trend as the result of patient data. So we conclude that the errors are trivial and wouldn't be of trouble when drawing the conclusions.

# 5 Conclusion

We conduct several experiments on 7 datasets using different preprocessing methods. From our work, changing parameters in preprocessing steps have little influence on the false positive rates. Our results imply that more preprocessing steps may not necessarily reduce false positives. Also, when designing the preprocessing pipeline, details about the scanning machine should be taken into consideration. This is a preliminary investigation on BOLD fMRI signal preprocessing by using classical statistical analysis method. Our future work will discuss how preprocessing may influence functional connectivity.

# 6 Acknowledgement

During the study, Professor Anders Eklund gives specific explanations and helpful advice through emails, which facilitates the study. We would like to thank professor Eklund for his help. Also, we thank Yueheng Wang and Huiwen Luo of Tsinghua University for discussions and suggestions during this study. We appreciate Tsinghua

Changgung Hospital (Beijing, China) for providing the patient data and guidance in this study. This work was supported by the Natural Science Foundation of China (NSFC-61171002, NSFC-60372023) and Tsinghua University Initiative Scientific Research Program (Grant No. 20131089382 and 20141081266).

No competing financial interests exist. No conflicts of interest.